\documentclass[notitlepage,aps,prl,reprint,superscriptaddress,longbibliography,groupedaddress]{revtex4-1}
\usepackage[utf8]{inputenc}

\usepackage{hyperref}
\usepackage{mathtools}
\usepackage{amsfonts}
\DeclarePairedDelimiter\bra{\langle}{\rvert}
\DeclarePairedDelimiter\ket{\lvert}{\rangle}
\usepackage{polski}

\begin{document}

\renewcommand{\figurename}{FIG.}
\renewcommand{\tablename}{TABLE}
\title{Generation of optical Gottesman-Kitaev-Preskill states with cavity QED}
\author{Jacob Hastrup}
\email{jhast@fysik.dtu.dk}
\author{Ulrik L. Andersen}
\affiliation{Center for Macroscopic Quantum States (bigQ), Department of Physics, Technical University of Denmark, Building 307, Fysikvej, 2800 Kgs. Lyngby, Denmark}

\begin{abstract}
Gottesman-Kitaev-Preskill (GKP) states are a central resource for fault-tolerant optical continuous-variable quantum computing. However, their realization in the optical domain remains to be demonstrated. Here we propose a method for preparing GKP states using a cavity QED system which can be realized in several platforms such as trapped atoms, quantum dots or diamond color centers. We then further combine the protocol with the previously proposed breeding protocol by Vasconcelos et al.\ to relax the demands on the quality of the QED system, finding that GKP states with more than 10 dB squeezing could be achieved in near-future experiments.
\end{abstract}


\maketitle

\section{Introduction}
Quantum error correction is an essential step towards building large scale quantum computers with realistic noisy components. In 2001, Gottesman, Kitaev and Preskill (GKP) proposed an error correction protocol in which each qubit is encoded into the continuous variables of an infinite dimensional bosonic mode \cite{gottesman2001encoding}. With this encoding, small errors such as displacements and losses of the bosonic mode \cite{albert2018performance,noh2018quantum} can be corrected using only Gaussian operations, thus providing an experimentally friendly and efficient framework, especially in the case where the bosonic mode is represented by an optical field. GKP error correction is particularly suitable in combination with optical cluster states \cite{menicucci2014fault,tzitrin2020progress,bourassa2021blueprint,larsen2021fault}, a field which have seen tremendous progress in recent years \cite{asavanant2019generation,larsen2019deterministic,larsen2020deterministic}. Additionally, GKP error correction can be used in long distance quantum communication schemes \cite{fukui2020all,rozpkedek2020quantum}, implementing quantum repeaters using only beamsplitters, homodyne detectors and GKP ancilla resource states \cite{walshe2020continuous}.

However, the encoded states themselves, denoted GKP states or grid states, are non-Gaussian and have proven extremely difficult to produce experimentally. Only in recent years have the states been produced in the motional mode of a trapped ion \cite{fluhmann2019encoding,de2020error} and in a microwave cavity field coupled to a superconducting circuit \cite{campagne2020quantum}. Crucially, they remain to be demonstrated in the optical domain, despite several proposed generation schemes \cite{pirandola2004constructing,pirandola2006continuous,vasconcelos2010all,weigand2018generating,su2019conversion,eaton2019non}. One promising approach is to interfere squeezed states on a multimode interferometer and project one output mode into an approximate GKP state by measuring the remaining modes with photon number resolving detectors \cite{su2019conversion,tzitrin2020progress}. Progress in high quality photon number resolving detectors could make this experiment feasible in the near future. However, the method is fundamentally probabilistic and thus needs multiplexing to be scalable, which imposes a large resource overhead. Furthermore, it is unclear how efficient and noise tolerant the protocol is for generating highly squeezed GKP states ($> 10$ dB squeezing), which are likely required to achieve fault-tolerance \cite{bourassa2021blueprint, larsen2021fault,tzitrin2021fault}. 

Another proposal is to build the GKP state using squeezed Schrödinger's cat states as the non-Gaussian element \cite{vasconcelos2010all,weigand2018generating}. The advantage of this approach is that it uses only beamsplitters and homodyne detectors, and that it can be made fully deterministic \cite{weigand2018generating}. However, it requires large amplitude cat states, which are challenging to produce in optics. Still, recently Hacker et al.\ demonstrated the experimental generation of optical cat states, by reflecting a light pulse off an optical cavity containing an atom \cite{hacker2019deterministic}. This method can in principle be used to generate cat states of arbitrary amplitude, although the method requires both high cooperativity and large escape efficiency which is experimentally challenging. 

In this work, inspired by the experimental progress reported in \cite{hacker2019deterministic}, we propose to use cavity quantum electrodynamics (QED) to generate approximate GKP states by iteratively reflecting squeezed states off a cavity containing a 3-level system. We thus extend the cat generation protocol of \cite{hacker2019deterministic} by inputting squeezed states, and by applying multiple interactions. We analyse the performance in systems with finite cooperativity and escape efficiency to determine the expected quality of the produced state with realistic devices. Furthermore, we combine the protocol with the cat breeding protocol of ref \cite{vasconcelos2010all}, which turns out to heavily relax the requirements on the quality of the cavity QED system. Finally, we propose a method to generate the input squeezed states also using the cavity QED system,  eliminating the need for a squeezed light source at the cavity QED resonance frequency.

\section{Preliminaries}
\subsection*{GKP states}
We describe the optical mode as a single bosonic mode with annihilation and creation operators $\hat{a}$ and $\hat{a}^\dagger$ and corresponding quadrature operators $\hat{x}=\frac{1}{\sqrt{2}}(\hat{a} + \hat{a}^\dagger)$ and $\hat{p}=\frac{1}{\sqrt{2}i}(\hat{a} - \hat{a}^\dagger)$ satisfying $[\hat{x},\hat{p}]=i$.\\
The aim of our work is to produce good approximate GKP states with a square lattice. In this work the relevant approximation is a finite superposition of squeezed states \cite{shukla2021squeezed}:
\begin{align}
    \ket{0_\textrm{GKP}} &\propto \sum_s \hat{D}\left(\sqrt{2\pi}s\right)\hat{S}(r)\ket{\textrm{vac}},  \nonumber \\
    \ket{1_\textrm{GKP}} &\propto \hat{D}\left(\sqrt{\pi/2}\right)\ket{0_\textrm{GKP}}, \label{eq:GKP}
\end{align}
where $\hat{D}(\alpha)=\exp(\alpha\hat{a}^\dagger-\alpha^*\hat{a})$ is the displacement operator and $\hat{S}(r)=\exp\left(\frac{1}{2}(r^*\hat{a}^2 - r\hat{a}^{\dagger2})\right)$ is the squeezing operator. The summation index, $s$, is over a finite number of integers around 0. GKP states have a periodic comb structure in both $x$ and $p$ quadratures with high quality GKP states consisting of highly squeezed peaks in both quadratures. Large squeezing in $x$ is achieved with large $r$ as is evident from Eq.\ \eqref{eq:GKP} while large squeezing in $p$ is achieved by including many terms in the sum. For a finite number of terms, the squeezing in $p$ can be further improved by weighing the superposition of Eq.\ \eqref{eq:GKP} such that terms further from the origin have less weight. 
In this work we quantify the quality of the produced GKP states by their amount of effective squeezing \cite{duivenvoorden2017single} in each quadrature, defined as:
\begin{align}
\Delta_x &= \sqrt{\frac{1}{2\pi}\ln\left(\frac{1}{|\langle\hat{D}(i\sqrt{2\pi})\rangle|^2}\right)}\\
\Delta_p& = \sqrt{\frac{1}{2\pi}\ln\left(\frac{1}{|\langle\hat{D}(\sqrt{2\pi})\rangle|^2}\right)}.
\end{align}
\label{eq:Delta}
The amount of squeezing is commonly denoted in dB as $\Delta_\textrm{dB}=-10\log_{10}(\Delta^2)$. For the approximate GKP state of Eq.\ \eqref{eq:GKP} one obtains $\Delta_x=e^{-r}$ while $\Delta_p$ depends on the number of terms, e.g. $\Delta_p = (6.6,10.4,13.7)$dB for $(2,4,8)$ terms respectively \cite{hastrup2021measurement}.

\subsection*{Cavity QED system}

\begin{figure}
    \centering
    \includegraphics{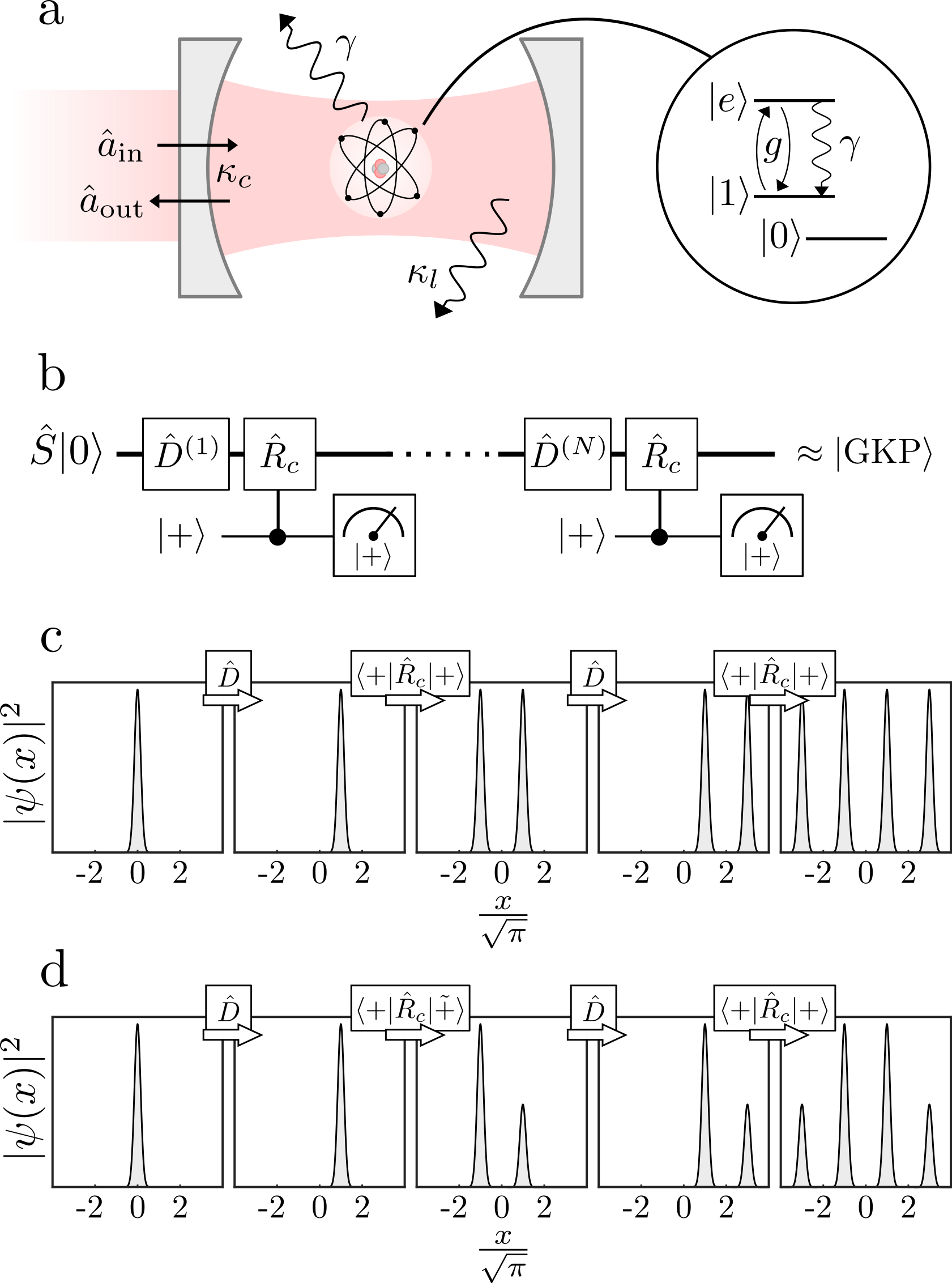}
    \caption{(a): Cavity QED system consisting of a cavity containing a 3-level system in which two levels resonantly couple to the cavity field. The cavity couples to a free-space field with rate $\kappa_c$ and to undesired scattering and loss modes with rate $\kappa_l$. Additionally, the excited state of the 3-level system can spontaneously decay trough modes different from the cavity mode with rate $\gamma$. Ideally, the cavity imprints a controlled rotation, $\hat{R}_c$ (Eq.\ \eqref{eq:Rc}), on the reflection of an incoming mode. (b) Circuit diagram for the GKP state generation protocol. (c) Repeated applications of displacements and controlled rotations evolves an initial squeezed vacuum state into an approximate GKP state. (d) Preparing the atom in an unequal superposition, $\ket{\tilde{+}}$, in the second to last step allows for the final state to have a two-level amplitude weighting of the squeezed peaks.}
    \label{fig:concept}
\end{figure}

Since GKP states are non-Gaussian we require a non-Gaussian element to generate them. In this work, we propose to use a cavity QED system as the central and only non-Gaussian element. In particular, we consider the reflection of an incoming optical field onto a single-mode cavity containing a 3-level system, as depicted in Fig.\ \ref{fig:concept}a. The 3-level system consists of two low energy states, $\ket{0}$ and $\ket{1}$, and one high energy state, $\ket{e}$, which can be optically excited from the state $\ket{1}$ through a Jaynes-Cumming Hamiltonian with coupling strength $g$. In this paper we denote this 3-level system as an ``atom'', e.g. as the one used in the experiment of \cite{hacker2019deterministic}. However, this atom could also be artificial such as a charged quantum dot \cite{lodahl2015interfacing,najer2019gated,hu2008giant} with the states $\ket{0}$ and $\ket{1}$ denoting spin states and $\ket{e}$ denoting a charged exciton state, or it could be a diamond color center \cite{janitz2020cavity,bhaskar2020experimental}, such as the nitrogen-vacancy center or silicon-vacancy center, where the states $\ket{0}$ and $\ket{1}$ are represented by spin ground states and $\ket{e}$ is an excited spin state. 
The cavity resonance frequency and the frequency of the incoming field are equal and tuned to match the $\ket{1}\leftrightarrow \ket{e}$ transition. To couple light into and out of the cavity, one end of the cavity is constructed with a slightly transparent mirror with a coupling rate $\kappa_c$ to an external free-space field. With the atom prepared in the $(\ket{0},\ket{1})$ subspace, an optical field mode reflected on the cavity ideally experiences a controlled phase rotation, $\hat{R}_c$, depending on the state of the atom \cite{duan2004scalable,hacker2019deterministic}:
\begin{equation}
     \hat{R}_c \equiv e^{i\pi\hat{n}}\otimes \ket{0}\bra{0} + \hat{\mathbb{I}}\otimes \ket{1}\bra{1}  \label{eq:Rc}
\end{equation}

 If the system is initially in the state $\ket{+} = (\ket{0} + \ket{1})/\sqrt{2}$, an incoming optical state, $\ket{\psi}$, evolves as:
 \begin{equation}
     \hat{R}_c\ket{\psi} \otimes \ket{+} = (e^{i\pi\hat{n}}\ket{\psi}\otimes \ket{0} + \ket{\psi}\otimes \ket{1})/\sqrt{2}.
 \end{equation}
Subsequently measuring the system in state $\ket{+}$ yields:
\begin{equation}
    \bra{+}\hat{R}_c\ket{\psi} \otimes \ket{+} = (e^{i\pi\hat{n}}\ket{\psi} + \ket{\psi})/\sqrt{2}.
\end{equation}
For example, for a coherent state input we obtain a Shrödinger's cat state, as was recently experimentally demonstrated \cite{hacker2019deterministic}. 

Realistic systems, however, are limited by losses and scattering into unwanted modes at rate $\kappa_l$, as well as spontaneous decay of the excited state of the atom through modes different than the cavity mode at rate $\gamma$. In the Supplementary Material we describe how to model these imperfections. The imperfections are conveniently described by the cooperativity, 
\begin{equation}
    C = \frac{g^2}{2\gamma\kappa},
\end{equation}
and escape efficiency
\begin{equation}
   \eta = \frac{\kappa_c}{\kappa},
\end{equation}
where $\kappa =\kappa_l+\kappa_c$ is the total cavity loss. Both $C$ and $\eta$ should preferably be as large as possible. However, there is a trade-off between the cooperativity and the escape efficiency. This is because the cooperativity can be increased by decreasing $\kappa_c$ while the escape efficiency is increased by increasing $\kappa_c$. Since we would like both high cooperativity and high escape efficiency one has to carefully tune the cavity coupling rate by engineering the cavity design. In the following we therefore quantify the system in terms of the internal cooperativity \cite{goto2019figure}, defined as
\begin{equation}
    C_0 = \frac{g^2}{2\kappa_l\gamma} = \frac{C}{1-\eta}. \label{eq:C0}
\end{equation}
Thus, for fixed $g$, $\kappa_l$ and $\gamma$, the internal cooperativity does not depend on the coupling rate $\kappa_c$. Note also that the internal cooperativity is always larger than the actual cooperativity, $C_0\geq C$. In the following analysis we numerically optimize $\kappa_c$ for each $C_0$ in order to optimize the effective squeezing of the output states.

\section{Results}
The idea of our proposed protocol is to repeatedly use the controlled rotation imposed by the cavity to generate an approximate GKP state, as illustrated in Fig. \ref{fig:concept}b and c. That is, inputting a displaced squeezed vacuum state we obtain a squeezed Schrödinger's cat state. Displacing and reflecting the output state on the cavity again further doubles the number of squeezed peaks in the output state and repeating this $N$ times yields a state of the form of Eq.\ \eqref{eq:GKP} with $2^N$ peaks. The displacement amplitude at the at $n$'th step is given by
\begin{equation}
    \hat{D}_n = \hat{D}\left(2^{n-1}\sqrt{\pi/2}\right). \label{eq:displacement}
\end{equation}
For a sufficiently squeezed input state the probability to obtain the measurement result $\ket{+}$ $N$ times is $1/2^N$. However, the first interaction can be made deterministic, by adding a feed-forward displacement operation since
\begin{align}
    \hat{D}\left(i\sqrt{\pi}/(2\sqrt{2})\right)\left(\hat{D}(\sqrt{2\pi}) - \hat{D}(-\sqrt{2\pi}) \right)\hat{S}(r)\ket{\textrm{vac}} \nonumber\\ 
    \approx \left(\hat{D}(\sqrt{2\pi}) + \hat{D}(-\sqrt{2\pi}) \right)\hat{S}(r)\ket{\textrm{vac}}
\end{align}
Thus a 4 peak state, which can yield up to 10.4 dB squeezing can be generated with probability 0.5, while an 8 peak state, yielding up to 13.7 dB squeezing, can be generated with probability 0.25. 

\begin{figure}
    \centering
    \includegraphics{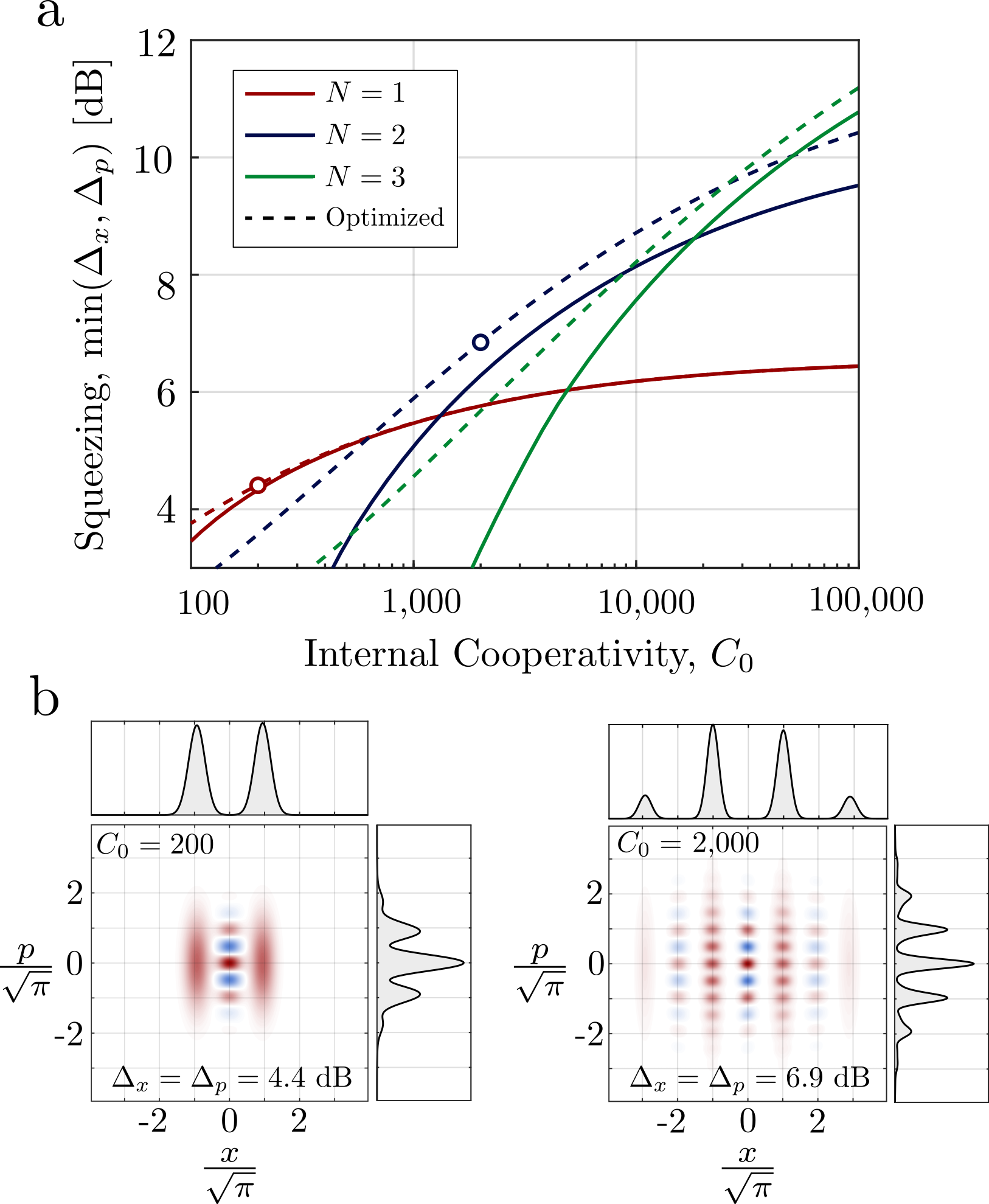}
    \caption{(a) Achievable amount of effective squeezing generated by the protocol outlined in Fig.\ \ref{fig:concept} with a noisy cavity, as a function of internal cooperativity, $C_0$ (Eq.\ \eqref{eq:C0}), optimizing the cavity output coupling rate and the input squeezing parameter. $N$ denotes the number of interactions with the cavity. The dashed lines further optimize over the displacement magnitudes and the initial state of the atom. (b) Wigner functions and quadrature distributions for the states generated with $C_0=200$ and $C_0 = 2000$ using $N=1$ and $N=2$ respectively, corresponding to open circles in (a).}
    \label{fig:results}
\end{figure}

The solid lines of fig.\ \ref{fig:results}a show the obtainable amount of squeezing using the protocol with finite-cooperativity systems. In addition to optimizing $\kappa_c$, we also numerically optimize the amount of input squeezing (See Supplementary Material for details on the input squeezing). The optimization is done by optimizing $\textrm{min}(\Delta_x,\Delta_p)$ such that we ensure effective squeezing in both quadratures. Additionally, we can slightly further improve the performance by slightly tuning the displacement amplitudes and the atomic superposition state (details in the Supplementary Information).

The dashed lines of Fig.\ \ref{fig:results}a show the result when implementing these two modifications. Note that for both the solid and dashed lines, there exists an optimal number of interactions, $N$, for each value of the internal cooperativity. This is because noise from the cavity adds up over multiple interactions, and thus at some point the noise added from the cavity outweighs the effect of increasing the number of peaks in the state. 
Fig.\ \ref{fig:results}b shows the Wigner functions and quadrature probability distributions of the achievable states with $C_0=200$ and $N=1$ (left) and $C_0=2000$ and $N=2$ (right). For $C_0=200$ the produced state is essentially a squeezed Schrödinger's cat state, but the quadrature distributions reveal the onset of the desired comb-like structure. For $C_0=2000$ we see a clear grid structure in the Wigner function and a narrowing of the peaks in the quadrature distribution.

As is evident from Fig.\ \ref{fig:results}, the protocol demands very high values of the internal cooperativity to produce high-squeezing grid states. This is due to the multiple interactions required with the noisy cavity, as well as the demanding simultaneous requirements of high cooperativity and high escape efficiency. 

To reduce the demands on the cavity QED system we propose to combine the protocol with the Schrödinger's cat state based breeding protocol of Ref.\ \cite{vasconcelos2010all}. In that protocol, we begin with a squeezed cat state of the form
\begin{equation}
    \ket{\textrm{sqcat}}=\left[\hat{D}\left(\sqrt{\pi}\sqrt{2}^{M-1}\right) + \hat{D}\left(-\sqrt{\pi}\sqrt{2}^{M-1}\right)\right]\hat{S}(r)\ket{\textrm{vac}}, \label{eq:cat}
\end{equation}
where $M$ is the number of iterations of the breeding protocol. Two such squeezed cat states are combined on a 50:50 beamsplitter and the $p$ quadrature of one mode is measured with a homodyne detector. Conditioned on the result $p=0$, the other mode is projected into an approximate GKP-like state. This protocol is then iterated, combining two such output states on another 50:50 beamsplitter and projecting one mode out with homodyne detection, etc. After $M$ iterations, the resulting output is an approximate GKP state of which $\Delta_p$ increases with the number of iterations and $\Delta_x$ equals the squeezing of the initial input cat states. One important feature of this breeding protocol is that homodyne detectors and beamsplitters can be implemented with near unity efficiency. Thus the experimental challenges are focused on producing high quality squeezed cat states. Note from Eq.\ \eqref{eq:cat} that the amplitude of the initial squeezed cat states depends on the number of iterations, $M$. Thus to achieve a highly squeezed approximate GKP state we require a large amplitude squeezed cat state, which is more sensitive to noise, such as loss, and thus experimentally more demanding. 

A deterministic version of this protocol was proposed in \cite{weigand2018generating}, by adding a feed-forward displacement to the final state. Furthermore, it was shown in \cite{weigand2018generating} that this deterministic approach on average generated GKP states with $\sim1$ dB more squeezing compared to the probabilistic approach. 

\begin{figure}[!h]
    \centering
    \includegraphics{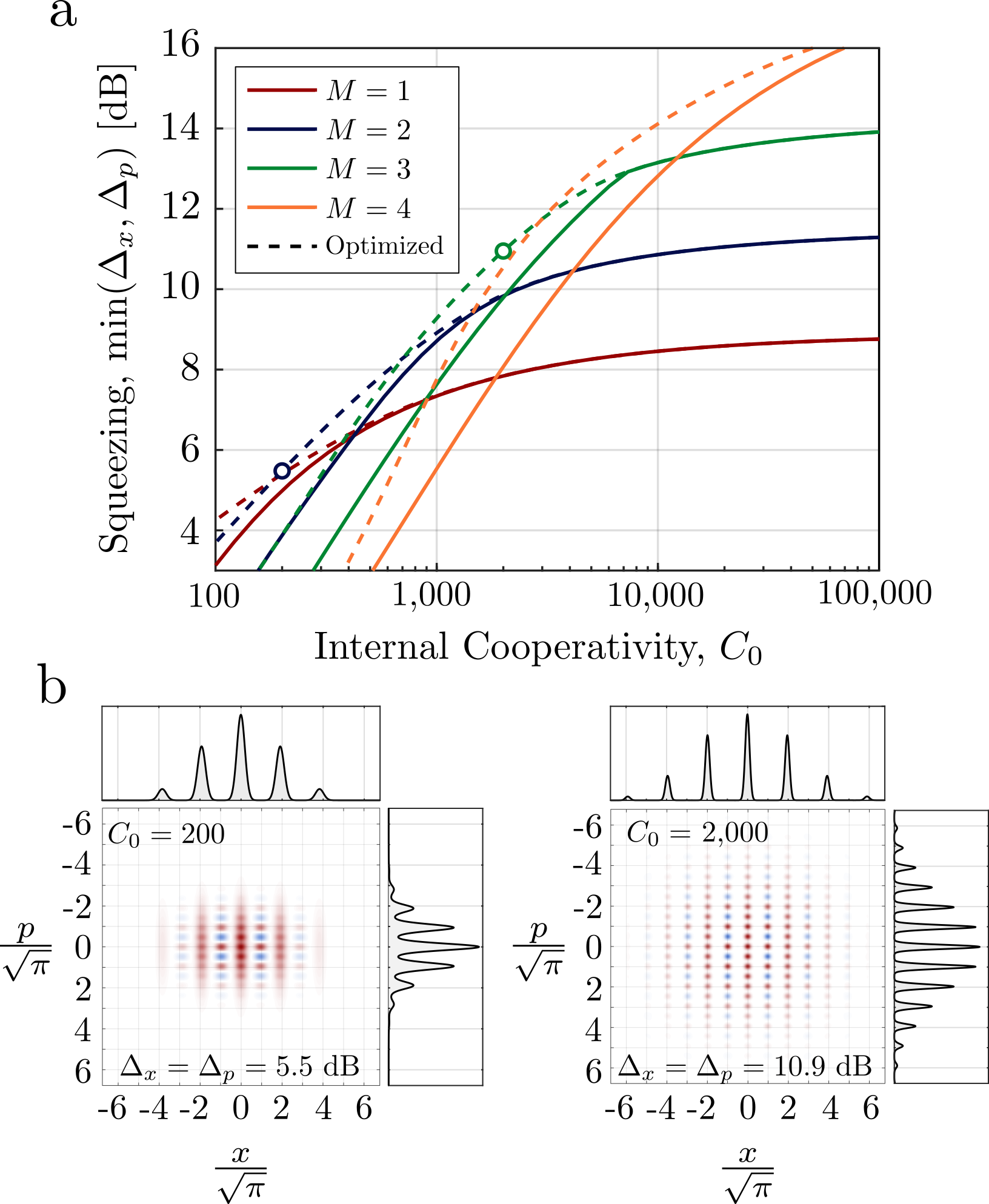}
    \caption{Achievable amount of effective squeezing obtained with the cat breeding method \cite{vasconcelos2010all}, using input squeezed cats generated with the cavity QED system. $M$ refers to the number of rounds of the breeding protocol. The dashed lines show the result when fine tuning the displacement amplitude to partly compensate losses in the cavity. (b) Wigner functions and quadrature distributions using $C_0=200$ and $C_0=2000$ with $M=2$ and $M=3$ respectively, corresponding to the points marked with open circles in (a).}
    \label{fig:breeding}
\end{figure}

Fig.\ \ref{fig:breeding}(a) shows the obtainable amount of effective squeezing generated with the breeding protocol, using squeezed cat states produced by a single reflection on the cavity. As with Fig.\ \ref{fig:results} we also optimize the displacement of the squeezed cat, with the results shown by the dashed lines. We see a substantial increase in the amount of achievable squeezing, reaching more than 10 dB for an internal cooperativity around $C_0=1300$, corresponding to a cooperativity of $C=25$ and escape efficiency $\eta=0.98$. Note that we have to generate $2^M$ squeezed cat states to breed each approximate GKP state. Even though each of these squeezed cats are generated under noisy conditions, they still breed into an approximate GKP state with more squeezing than what is possible solely using the cavity. Fig.\ \ref{fig:breeding}b shows Wigner functions and quadrature distributions of two example states generated with $C_0=200$ and $C_0=2000$ using $M=2$ and $M=3$ respectively. Comparing to Fig.\ \ref{fig:results}c we observe clear improvement in the quality of the produced states. 

The results presented in Fig.\ \ref{fig:breeding} are generated using the original probabilistic approach \cite{vasconcelos2010all}, as it allows efficient evaluation of the effective squeezing levels with mixed state inputs, which enables us to numerically optimize the cavity coupling rate and input squeezing levels. However, as mentioned, the protocol can be made fully deterministic following \cite{weigand2018generating}, with the added benefit of an expected slight increase in the squeezing levels.

Finally, we address the input squeezed light source. Ideally, one might want to use squeezed light generated from parametric down conversion, as this method can yield very high squeezing values. However, the wavelength and temporal mode profile of the squeezed light from such a source might not be straightforwardly compatible with a high cooperatively cavity QED system. In the Supplementary Material we therefore propose a method to generate squeezed states starting from a coherent state, using the cavity QED system.

\section{Conclusion}
We have presented a method for generating approximate GKP states using a cavity QED system as the central non-Gaussian element. The performance is in practice limited by the internal cooperativity of the systems. State of the art system have demonstrated internal cooperatives of up to 200 \cite{najer2019gated,bhaskar2020experimental} which could produce approximate GKP states with 4.4 dB squeezing, which can be improved to 5.5 dB through the breeding method of \cite{vasconcelos2010all}. However, improved cavity designs are rapidly being developed across multiple platforms, and designs with cooperativities exceeding 1000 have been proposed \cite{al2018cooperativity}, which could push the achievable amount of effective squeezing above 10 dB in the near future.


\section*{Acknowledgements}
This project was supported by the Danish National Research Foundation through the Center of Excellence for Macroscopic Quantum States (bigQ, DNRF0142).

\section{References}
\bibliography{References}


\clearpage
\begin{widetext}
\section*{Supplementary Material}
\section{S1: Realistic reflection channel}
Here we show how to model the realistic cavity described by finite cooperativity and escape efficiency. We wish to determine reflected field mode by a quantum channel described by Kraus operators as
\begin{equation}
    \rho \rightarrow \sum_{(m_l,m_\gamma)\in\mathbb{N}_0^2} \hat{K}_{m_l,m_\gamma}\rho \hat{K}_{m_l,m_\gamma}^\dagger, \label{eq:Kraus}
\end{equation}
where $\rho$ denotes the density matrix of the incoming optical field. As we shall show in the following, the Kraus operator $\hat{K}_{m_l,m_\gamma}$ corresponds to the event of losing $m_l$ photons to unwanted cavity losses and $m_\gamma$ photons via scattering of the atom. We use standard input-output theory to model the system \cite{gardiner1985input,walls2007quantum}. Thus the input and output fields, $\hat{a}_\textrm{in}$ and $\hat{a}_\textrm{out}$, are related to the cavity field, $\hat{a}_c$ in the Heisenberg picture as:
\begin{equation}
    \hat{a}_\textrm{out} = \sqrt{2\kappa_c}\hat{a}_c+\hat{a}_\textrm{in}. \label{eq:inputoutput}
\end{equation}
The quantum Langevin equation for the cavity field operator, including excess losses to the mode $\hat{a}_l$ is given by:
\begin{equation}
    \dot{\hat{a}}_c = -\frac{i}{\hbar}[\hat{a}_c,\hat{H}]-\sqrt{2\kappa_c}\hat{a}_\textrm{out} - \sqrt{2\kappa_l}\hat{a}_l + \kappa\hat{a}_c,
\end{equation}
where $\hat{H}$ is the cavity Hamiltonian and $\kappa = \kappa_l + \kappa_c$. We consider first the case of the atom in the $\ket{1},\ket{e}$-subspace. The cavity Hamiltonian is given by the Jaynes-Cummings Hamiltonian:
\begin{equation}
    \hat{H} = \hbar\omega_c\hat{a}_c^\dagger\hat{a}_c + \hbar\omega_a\hat{\sigma}_z/2 + \hbar g\left(\hat{a}_c\hat{\sigma}_+ + \hat{a}_c^\dagger \hat{\sigma}_-\right),
\end{equation}
where $\hat{\sigma}_z=-\ket{1}\bra{1}+\ket{e}\bra{e}$, $\hat{\sigma}_-=\ket{1}\bra{e}$ and $\hat{\sigma}_+=\ket{e}\bra{1}$ and $g$ is the coupling rate. Additionally, we consider the quantum Langevin equation for the operator $\hat{\sigma}_-$, including atomic decay into modes different from the cavity mode, denoted $\hat{a}_\gamma$:
\begin{equation}
    \dot{\hat{\sigma}}_- = -\frac{i}{\hbar}[\hat{\sigma}_-,\hat{H}] + \hat{\sigma}_z(-\gamma\hat{\sigma}_- + \sqrt{2\gamma}\hat{a}_\gamma).
\end{equation}
Inserting $\hat{H}$ we get:
\begin{align}
    \dot{\hat{a}}_c&=-i\omega_c\hat{a}_c - ig\hat{\sigma}_-  -\sqrt{2\kappa_c}\hat{a}_\textrm{out} - \sqrt{2\kappa_l}\hat{a}_l + \kappa\hat{a}_c\\
    \dot{\hat{\sigma}}_-&=-i\omega_a\hat{\sigma}_- + ig\hat{a}_c\hat{\sigma}_z + \hat{\sigma}_z(-\gamma\hat{\sigma}_- + \sqrt{2\gamma}\hat{a}_\gamma).
\end{align}
These are solved for the cavity field in the frequency domain at resonance, assuming weak excitation of the atom such that $\langle \hat{\sigma}_z\rangle = -1$:
\begin{equation}
    \hat{a}_c(\omega) = \frac{\sqrt{2\kappa_c}\hat{a}_\textrm{out}(\omega) + \sqrt{2\kappa_l}\hat{a}_l(\omega)  + \frac{ig\sqrt{2\gamma}}{\gamma - i\Delta_a}\hat{a}_\gamma(\omega)}{\kappa - i\Delta_c + \frac{g^2}{\gamma - i\Delta_a}}
\end{equation}
where $\Delta_c = \omega_c - \omega$ and $\Delta_a = \omega_a - \omega$. 
Inserting into the input-output relation, \eqref{eq:inputoutput}:

\begin{align}
    \hat{a}_\textrm{in}(\omega) = \frac{(\kappa - i\Delta_c + \frac{g^2}{\gamma - i\Delta_a} - 2\kappa_c)\hat{a}_\textrm{out}(\omega)  - 2\sqrt{\kappa_l\kappa_c}\hat{a}_l(\omega) - \frac{i2g\sqrt{\gamma\kappa_c}}{\gamma - i\Delta_a }\hat{a}_\gamma(\omega)}{\kappa - i\Delta_c + \frac{g^2}{\gamma - i\Delta_a}} = r_1\hat{a}_\textrm{out}(\omega) + t_1\hat{a}_l(\omega) + \Gamma \hat{a}_\gamma(\omega)
\end{align}
At resonance, $\Delta_c = 0$ and $\Delta_a = 0$, the coefficients can be written in terms of the cooperativity, $C = g^2/(2\gamma\kappa)$, and escape efficiency, $\eta=\kappa_c/\kappa$ as:
\begin{align}
    r_1 &= \frac{2C+1-2\eta}{2C+1} \\
    t_1&= -\frac{2\sqrt{\eta(1-\eta)}}{2C + 1} \\
    \Gamma &=  -i\frac{2\sqrt{2\eta C}}{2C + 1}.
\end{align}
In the case of the atom in state $\ket{0}$ the relevant coefficients are obtained by setting $g=0$ (corresponding to $C=0$). This gives:
\begin{align}
    r_0 &= 1 - 2\eta \\
    t_0 &= -2\sqrt{\eta(1 - \eta)},
\end{align}
and no scattering from the atom. In total, the input field transforms as:
\begin{align}
    \hat{a}_\textrm{in} \rightarrow  (r_0\hat{a}_\textrm{out} + t_0\hat{a}_l)\otimes\ket{0}\bra{0}  +  (r_1\hat{a}_\textrm{out} + t_1\hat{a}_l + \Gamma\hat{a}_\gamma)\otimes\ket{1}\bra{1} \label{eq:transformation}
\end{align}
To find the corresponding Kraus operators we consider the transformation of an arbitrary pure input state,
\begin{equation}
\ket{\psi} = \sum_{n=0}^{\infty} c_n\ket{n}_{\textrm{in}} = \sum_{n=0}^{\infty}c_n \frac{(\hat{a}_\textrm{in}^\dagger)^n}{\sqrt{n!}}\ket{\textrm{0}}_{\textrm{in}}.
\end{equation}
Inserting Eq.\ \eqref{eq:transformation}:
\begin{align}
    \ket{\psi}\rightarrow & \sum_n\frac{c_n}{\sqrt{n!}}\Big[(r_0\hat{a}_\textrm{out}^\dagger + t_0\hat{a}_l^\dagger)\otimes\ket{0}\bra{0} +  (r_1\hat{a}_\textrm{out}^\dagger + t_1\hat{a}_l^\dagger + \Gamma\hat{a}_\gamma^\dagger)\otimes\ket{1}\bra{1}\Big]^n\ket{\textrm{0}}_{\textrm{out}}\ket{\textrm{0}}_l\ket{\textrm{0}}_{\gamma} \\
    =\, & \Big[\sum_n\frac{c_n}{\sqrt{n!}}(r_0\hat{a}_\textrm{out}^\dagger + t_0\hat{a}_l^\dagger)^n\Big]\ket{\textrm{0}}_{\textrm{out}}\ket{\textrm{0}}_{l}\ket{\textrm{0}}_{\gamma}\otimes\ket{0}\bra{0} 
    +  \Big[\sum_n\frac{c_n}{\sqrt{n!}}(r_1\hat{a}_\textrm{out}^\dagger + t_1\hat{a}_l^\dagger + \Gamma\hat{a}_\gamma^\dagger)^n\Big]\ket{\textrm{0}}_{\textrm{out}}\ket{\textrm{0}}_{l}\ket{\textrm{0}}_{\gamma}\otimes\ket{1}\bra{1} \label{eq:S1}
\end{align}
For simplicity we consider now the term containing $\ket{1}\bra{1}$. The $\ket{0}\bra{0}$ is expanded in a similar fashion:
\begin{align}
&\sum_n\frac{c_n}{\sqrt{n!}}(r_1\hat{a}_\textrm{out}^\dagger + t_1\hat{a}_l^\dagger + \Gamma\hat{a}_\gamma^\dagger)^n\ket{\textrm{0}}_{\textrm{out}}\ket{\textrm{0}}_{l}\ket{\textrm{0}}_{\gamma}\nonumber \\
= &\,\sum_n\frac{c_n}{\sqrt{n!}}\sum_{m=0}^n\begin{pmatrix}n\\m \end{pmatrix} (r_1\hat{a}_\textrm{out}^\dagger)^{n-m} (t_1\hat{a}^\dagger + \Gamma\hat{a}_\gamma^\dagger)^m\ket{\textrm{0}}_{\textrm{out}}\ket{\textrm{0}}_{l}\ket{\textrm{0}}_{\gamma}\\
=&\,\sum_{n=0}^\infty\frac{c_n}{\sqrt{n!}}\sum_{m=0}^n\begin{pmatrix}n\\m \end{pmatrix} (r_1\hat{a}_\textrm{out}^\dagger)^{n-m}\sum_{m_\gamma=0}^m\begin{pmatrix}m \\ m_\gamma\end{pmatrix} (t_1\hat{a}_l^\dagger)^{m-m_\gamma} (\Gamma\hat{a}_\gamma^\dagger)^{m_\gamma}\ket{\textrm{0}}_{\textrm{out}}\ket{\textrm{0}}_{l}\ket{\textrm{0}}_{\gamma}.
\end{align}
Reordering the summations as $\sum_{n=0}^\infty\sum_{m=0}^n\sum_{m_\gamma=0}^m = \sum_{m=0}^\infty\sum_{m_\gamma=0}^m\sum_{n=m}^\infty$ and applying the creation operators:
\begin{align}
    =&\sum_{m=0}^\infty\sum_{m_\gamma=0}^m\sum_{n=m}^\infty \frac{c_n}{\sqrt{n!}}\begin{pmatrix}n\\m\end{pmatrix}\begin{pmatrix}m \\ m _\gamma\end{pmatrix}r_1^{n-m}t_1^{m-m_\gamma}\Gamma^{m_\gamma} \\
    &\times \sqrt{(n-m)!}\sqrt{(m-m_\gamma)!}\sqrt{m_\gamma!}\ket{n-m}_\textrm{out}\ket{m-m_\gamma}_l\ket{m_\gamma}_\gamma
\end{align}
When tracing out the lossy modes, we obtain an incoherent mixture of the terms in the inner sum. Each of these terms corresponds to losing $m_\gamma$ photons from scattering of the atom and $m_l=m-m_\gamma$ photons to excess cavity losses. Looking at one of these terms:
\begin{align}
    &\sum_{n=m_l + m_\gamma}^\infty \frac{c_n}{\sqrt{n!}}\begin{pmatrix}n \\ m_l + m_\gamma \end{pmatrix}\begin{pmatrix} m_l + m_\gamma \\ m_\gamma \end{pmatrix} r_1^{n-m_l - m_\gamma}t_1^{m_l}\Gamma^{m_\gamma}\sqrt{m_l!}\sqrt{m_\gamma!} \sqrt{(n-m_l-m_\gamma)!}\ket{n-m_l-m_\gamma} \\
    =&\, \sum_{n=m_l+m_\gamma}^{\infty}\frac{c_n}{\sqrt{n}!}\left(\frac{t_1}{r_1}\right)^{m_l}\left(\frac{\Gamma}{r_1}\right)^{m_\gamma}\frac{1}{\sqrt{m_l!m_\gamma!}} \hat{a}_\textrm{out}^{m_l + m_\gamma}r_1^{\hat{n}}\ket{n} \\
    =&\, \left(\frac{t_1}{r_1}\right)^{m_l}\left(\frac{\Gamma}{r_1}\right)^{m_\gamma}\frac{1}{\sqrt{m_l!m_\gamma!}} \hat{a}_\textrm{out}^{m_l + m_\gamma}r_1^{\hat{n}}\ket{\psi}
\end{align}
Similarly, when including also the $\ket{0}\bra{0}$ term in Eq.\ \eqref{eq:S1}, the term corresponding to losing ($m_l$,$m_\gamma$) photons is:
\begin{equation}
    \left[\delta_{m_\gamma,0}\left(\frac{t_0}{r_0}\right)^{m_l}\frac{\hat{a}_\textrm{out}^{m_l}}{\sqrt{m_l!}}r_0^{\hat{n}}\otimes \ket{0}\bra{0}+\left(\frac{t_1}{r_1}\right)^{m_l}\left(\frac{\Gamma}{r_1}\right)^{m_\gamma}\frac{\hat{a}_\textrm{out}^{m_l + m_\gamma}}{\sqrt{m_l!m_\gamma!}} r_1^{\hat{n}}\otimes\ket{1}\bra{1}\right]\ket{\psi}\equiv \hat{K}_{m_l,m_\gamma}\ket{\psi}.
\end{equation}
From this we identify the Kraus operator corresponding to the loss of ($m_l$,$m_\gamma$) photons:
\begin{equation}
    \hat{K}_{m_l,m_\gamma} = \delta_{m_\gamma,0}\left(\frac{t_0}{r_0}\right)^{m_l}\frac{\hat{a}^{m_l}}{\sqrt{m_l!}}r_0^{\hat{n}}\otimes \ket{0}\bra{0}+\left(\frac{t_1}{r_1}\right)^{m_l}\left(\frac{\Gamma}{r_1}\right)^{m_\gamma}\frac{\hat{a}^{m_l + m_\gamma}}{\sqrt{m_l!m_\gamma!}} r_1^{\hat{n}}\otimes\ket{1}\bra{1}.
\end{equation}


\section*{S2: Squeezing in the limit $\eta\rightarrow 1$ and $C\rightarrow\infty$}
\begin{figure}
    \centering
    \includegraphics{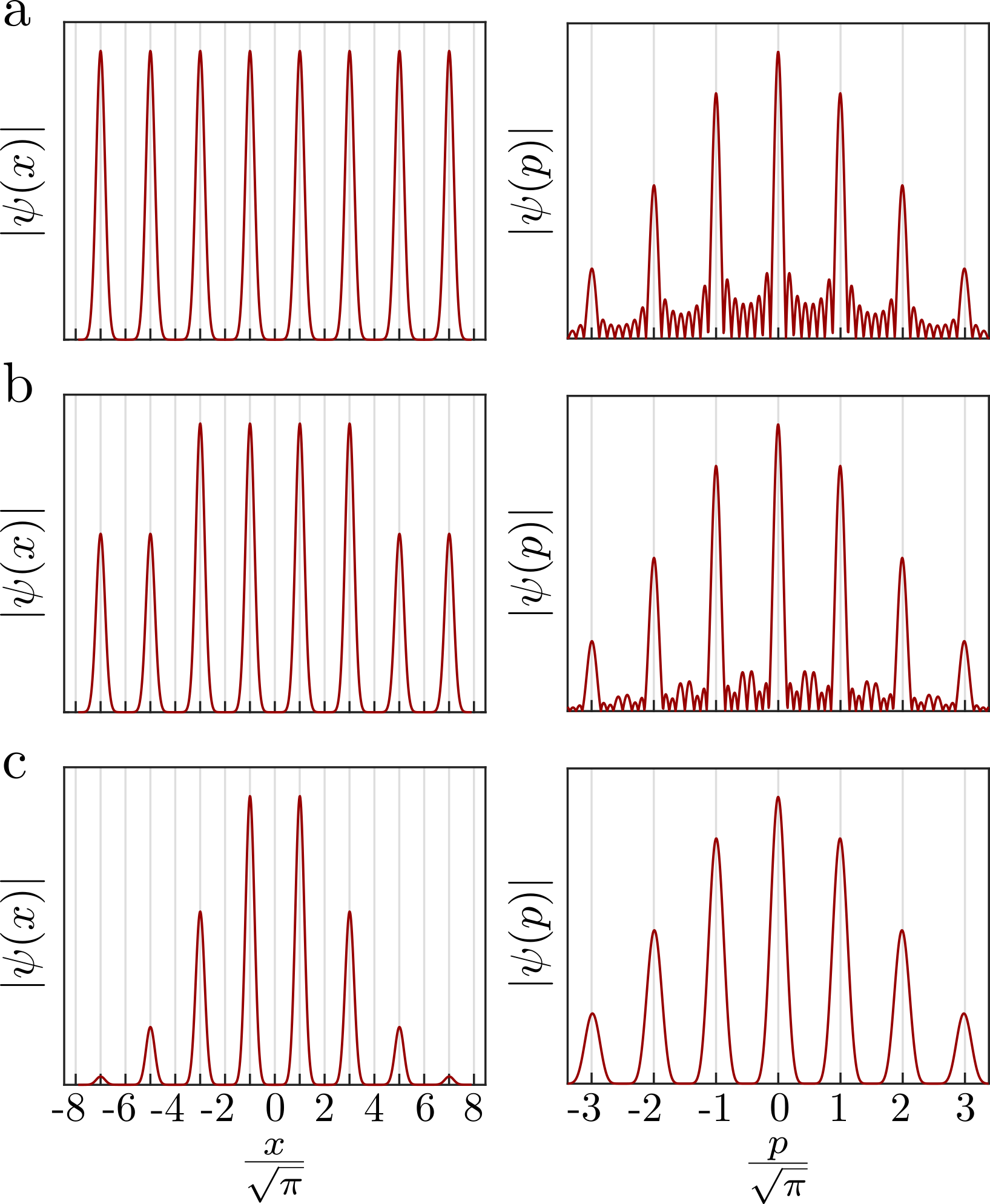}
    \caption{$x$ and $p$ quadrature distributions for different approximate GKP states, each composed of $N_\textrm{peaks}=8$ peaks of 10 dB squeezing with different peaks weightings. (a) Equal weighting. (b) Two-level weighting. (c) Binomial weighting.  }
    \label{fig:Distributions}
\end{figure}

Here we calculate the effective squeezing levels of ideal states generated by the protocol. These can be written in the form
\begin{equation}
    \ket{\psi_\textrm{GKP}}=\sum_s c_s\hat{D}\left(\sqrt{\pi/2}s\right)\hat{S}(r)\ket{\textrm{vac}}. \label{eq:GKPgeneral}
\end{equation}
For logical 0 states, $c_s$ is zero for odd $s$, while for logical 1 states, $c_s$ is zeros for even $s$. For sufficiently large $r$ such that neighbouring states are non overlapping, normalization is achieved by $\sum_s |c_s|^2=1$. The expectations value used to calculate the the effective squeezing level is given by
\begin{align}
    \langle\hat{D}\left(i\sqrt{2\pi}\right)\rangle &= e^{-\pi e^{-2r}}, \\
    \langle\hat{D}\left(\sqrt{2\pi}\right)\rangle &= \sum c_s^* c_{s+2},\label{eq:expectation}
\end{align}
again assuming negligible overlap between neighbouring states. Thus the effective squeezing level in the $x$ quadrature depends only on $r$, while the effective squeezing in the $p$ quadrature is independent of $r$ and depends instead only on the distribution of the peaks. 

\subsubsection*{Equal weighting}
The states generated directly with the cavity, i.e. by the circuit of Fig.\ \ref{fig:concept}b after $N$ steps have $N_\textrm{peaks} = 2^N$ non-zero coefficients of equal amplitude $1/\sqrt{N_\textrm{peaks}}$. Fig.\ \ref{fig:Distributions}a shows the quadrature distributions of this state with $8$ peaks. For $N_\textrm{peaks}$ peaks Eq.\ \eqref{eq:expectation} yields:
\begin{equation}
    \langle\hat{D}\left(\sqrt{2\pi}\right)\rangle = \frac{N_\textrm{peaks}-1}{N_\textrm{peaks}}.\label{eq:constant}
\end{equation}
For $N_\textrm{peaks}=2$, 4 or 8 we then get $\Delta_p=6.6$dB, $10.4$dB or $13.7$ dB respectively. In the limit of many peaks, i.e. large $N_\textrm{peaks}$ we get $\Delta_p\approx \sqrt{1/(\pi N_\textrm{peaks})}$ and thus doubling the number of peaks increases the amount of squeezing with $3$ dB. 

\subsubsection*{Two-level weighting}
As mentioned in the main text, the effective squeezing can increased by imposing an envelope over the coefficients $c_s$. In our protocol, we can create a two-level envelope by preparing the atom in an uneven superposition, $a\ket{0} + b\ket{1}$ with $|a|^2 + |b|^2=1$, in the second to last interaction. In the resulting normalized output GKP state, the innermost half of the peaks have amplitude $\sqrt{2}a/\sqrt{N_\textrm{peaks}}$ while the outermost half have amplitude $\sqrt{2}b/\sqrt{N_\textrm{peaks}}$, as shown in Fig.\ \ref{fig:Distributions}b. In this case we get from Eq.\ \eqref{eq:expectation}:
\begin{equation}
    \langle\hat{D}\left(\sqrt{2\pi}\right)\rangle = \frac{N_\textrm{peaks}-(4|b|^2 + 2|a|^2-4\textrm{Re}[a^*b])}{N_\textrm{peaks}}.
    \end{equation}
This expression is optimized for $a=\sqrt{1/2 + 1/\sqrt{20}}$ and $b=\sqrt{1/2 - 1/\sqrt{20}}$, in which case we get
\begin{equation}
    \langle\hat{D}\left(\sqrt{2\pi}\right)\rangle = \frac{N_\textrm{peaks} - \overbrace{(3-\sqrt{5})}^{\sim0.76}}{N_\textrm{peaks}}.
\end{equation}
In the limit of large $N_\textrm{peaks}$ we get $\Delta_p\approx \sqrt{3-\sqrt{5}}\sqrt{1/(\pi N_\textrm{peaks})}$ corresponding to $1.2$ dB more squeezing compared to the equal-amplitude distribution. Comparing Fig.\ \ref{fig:Distributions}a and b we see that imposing the two-level weighting of the peaks helps to reduce the noise of the $p$-distribution.

\subsubsection*{Binomial weighting}
The states generated with the cat-breeding protocol also take the form of Eq.\ \eqref{eq:GKPgeneral}, but with $N_\textrm{peaks}=2^{M}+1$ peaks forming a binomial distribution of the coefficients \cite{vasconcelos2010all}. Thus the expectation value from Eq.\ \eqref{eq:expectation} is:
\begin{align}
    \langle\hat{D}\left(\sqrt{2\pi}\right)\rangle& = \frac{\sum_{k=0}^{N_\textrm{peaks}-2}\begin{pmatrix}N_\textrm{peaks}-1\\k \end{pmatrix} \begin{pmatrix}N_\textrm{peaks}-1\\k+1 \end{pmatrix}}{\sum_{k=0}^{N_\textrm{peaks}-2}\begin{pmatrix}N_\textrm{peaks}-1\\k \end{pmatrix}^2}\\
    &=\frac{\begin{pmatrix} 2(N_\textrm{peaks}-2) \\ N_\textrm{peaks}-1\end{pmatrix}}{\begin{pmatrix} 2(N_\textrm{peaks}-1) \\ N_\textrm{peaks}-1\end{pmatrix}} \\
    &=\frac{N_\textrm{peaks}-1}{N_\textrm{peaks}},
\end{align}
where the 2nd equality follows from Vandermonde's identity in both the numerator and denominator. Coincidentally, this expectation value is the same as for the equal weighting, Eq.\ \eqref{eq:constant}, despite the quadrature distributions being quite different. Fig.\ \ref{fig:Distributions}c shows the binomial weighting with 8 peaks. Comparing to the equal weighting, Fig.\ \ref{fig:Distributions}a, the binomial distribution greatly reduces the $p$-quadrature noise floor, but at the cost of broadening the peaks. This trade-off happens to be such that $\langle\hat{D}\left(\sqrt{2\pi}\right)\rangle$ remains unchanged. Note that this does not necessarily imply that the two states perform identically in practice, as this will depend on the specifics of the error correction protocol.


\section*{S3: Input squeezing}
\begin{figure}
    \centering
    \includegraphics{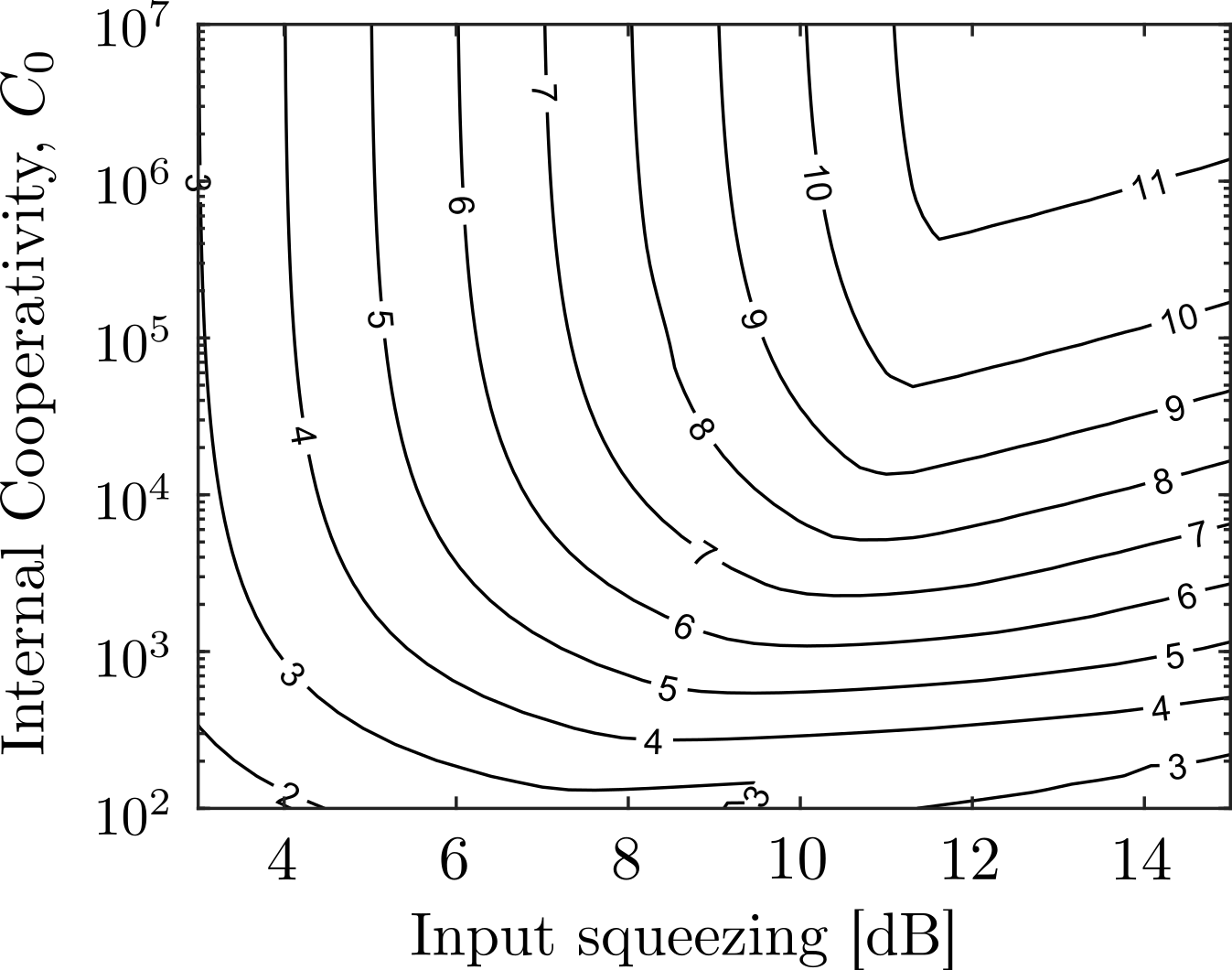}
    \caption{Effective squeezing in the least squeezed quadrature, $\textrm{min}(\Delta_x,\Delta_p)$, in dB of the states generated with $N=2$ as a function of input squeezing and internal cooperativity.}
    \label{fig:InputSqueezing}
\end{figure}

Fig.\ \ref{fig:InputSqueezing} shows the achievable effective squeezing with $N=2$ as a function of internal cooperativity and input squeezing, numerically optimizing the coupling rate and displacement amplitudes at each point. For a given internal cooperativity there exists an optimum input squeezing level. This is because a heavily $x$-squeezed input state has a larger envelope in the $p$-quadrature. As the state experiences losses during generation, peaks at large $|p|$ experience a shift towards 0, thus shifting relative to the GKP lattice which degrades the effective squeezing level. 
In general, the achievable output squeezing is a few dB lower than the optimum input squeezing. This is because the effective squeezing in the $x$-quadrature is completely determined by the input squeezing. As the input state experiences noise from the cavity, the squeezing level in the $x$-quadrature is thus reduced by a few dB in the produced state, compared to the squeezing of the input state.


\section*{S4: Displacement amplitudes}
\begin{figure}
    \centering
    \includegraphics{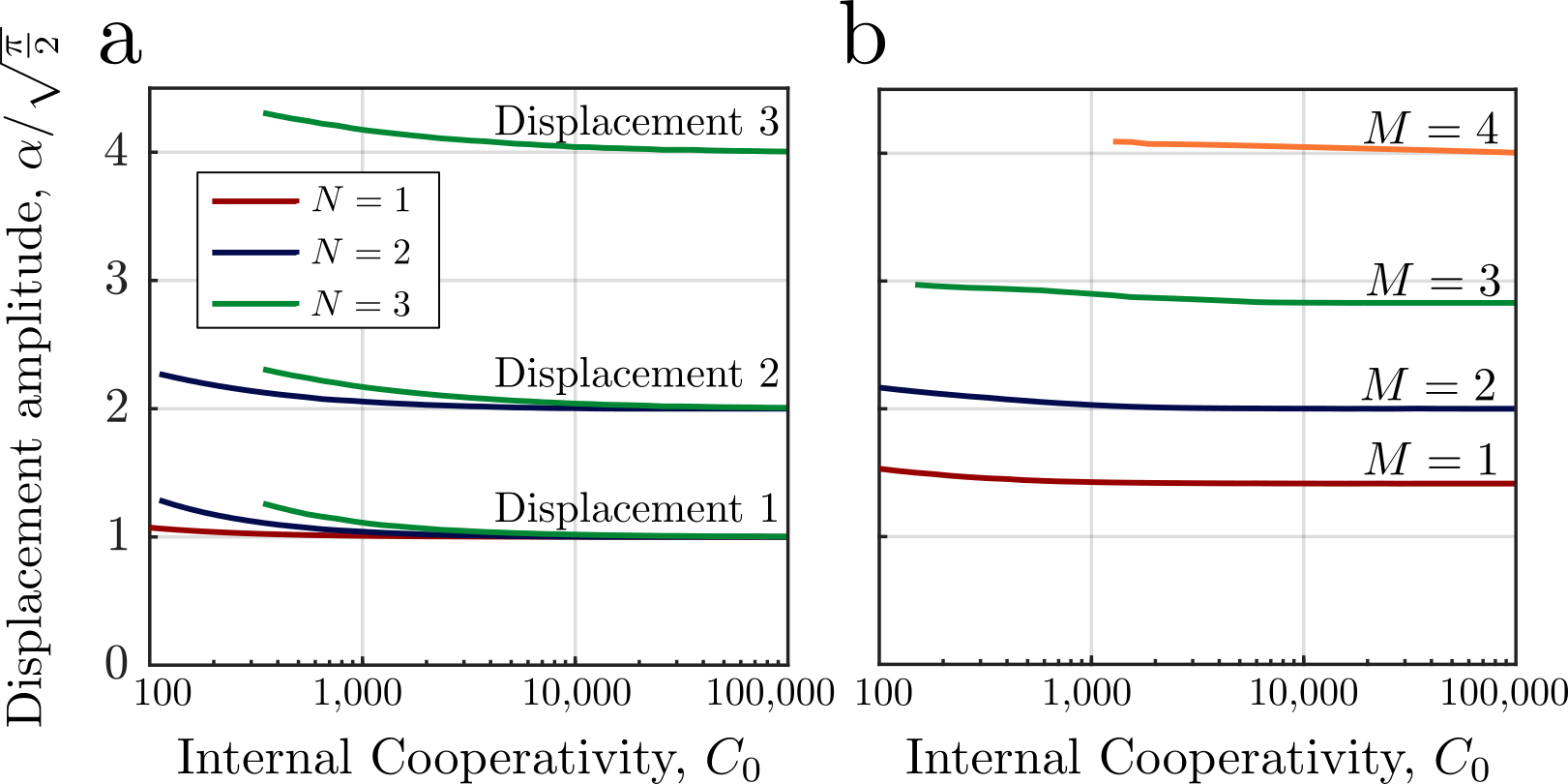}
    \caption{Numerically optimized displacement amplitudes as a function of internal cooperativity.}
    \label{fig:Scale}
\end{figure}
Fig. \ref{fig:Scale} shows the numerically optimized displacement amplitudes. When the internal cooperativity is large, the optimal amplitudes converge to that of Eq.\ \eqref{eq:displacement} for the cavity-only protocol (a) and $\sqrt{\pi}\sqrt{2}^{M-1}$ for the cat breeding protocol (b), as expected. However, when the internal cooperativity is smaller, we find that the optimal displacement is slightly larger than for the perfect cavity. The slightly larger displacement partly compensates for the loss induced by the cavity. Note that the compensation only works because the noise of cavity reflection channel is not described by pure loss, but by the more complicated channel of Eq.\ \eqref{eq:Kraus}. The input state thus effectively experience more loss than the coherence between the generated peaks \cite{hacker2019deterministic}, which enable a slight compensation to re-position the peaks on the GKP lattice in the $x$-quadrature without causing unwanted shrinking of the lattice in the $p$-quadrature. 


\begin{figure}
    \centering
    \includegraphics{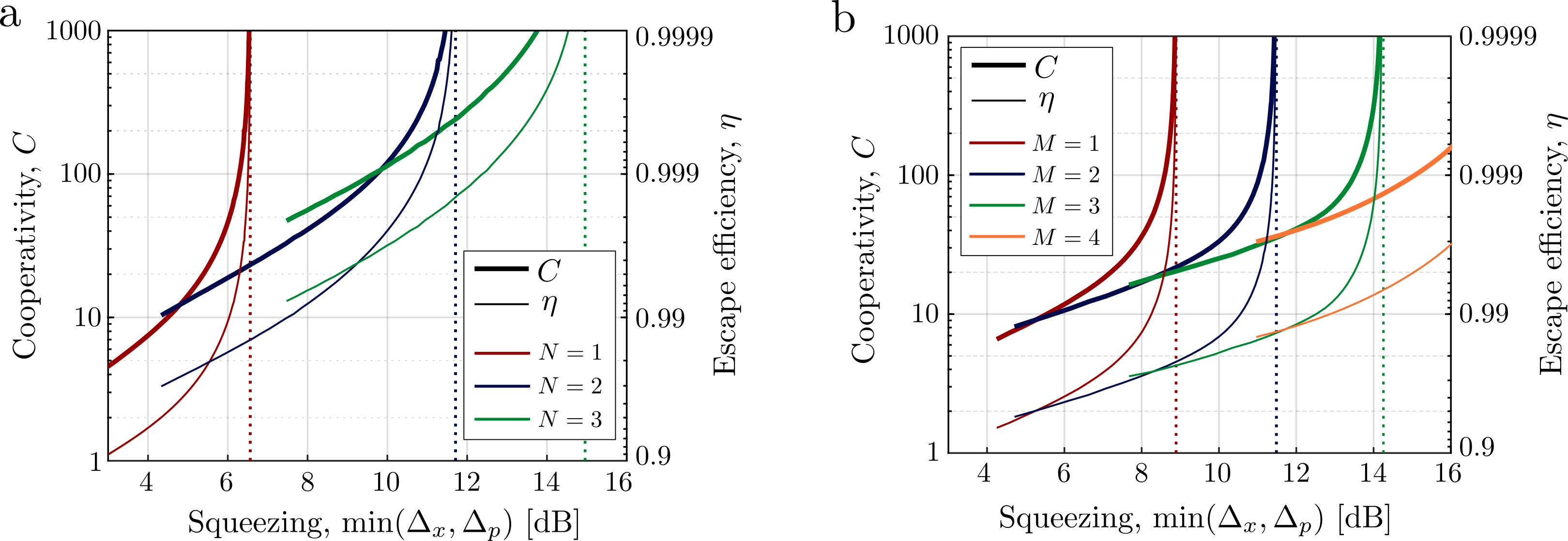}
    \caption{Optimal cooperativity (left axis, thick lines) and escape efficiency (right axis, thin lines) as a function of the achievable effective squeezing using only the cavity (a) and with the cat-breeding method (b). The vertical lines shows the achievable amount of effective squeezing in the limit of $\eta\rightarrow1$ and $C\rightarrow \infty$. }
    \label{fig:Cooperativity}
\end{figure}

\section*{S5: Required cooperativity and escape efficiency}
As discussed in the main text one should optimize the cavity coupling rate to ensure simultaneous high escape efficiency and cooperativity. Fig.\ \ref{fig:Cooperativity} shows the resulting optimized values of the cooperativity and escape efficiency as a function of the achievable amount of squeezing with a fixed internal cooperativity. For example, 10 dB squeezing can be obtained using $N=3$ interactions with a system with $C=110$ and escape efficiency $\eta = 0.997$ using the cavity only or $C=25$ and $\eta = 0.98$ with $M=3$ steps of the cat-breeding protocol.


\section*{S6: Calculation of effective squeezing parameters}
To perform simulations of the reflection channel, we represent our state numerically in the Fock basis with a max photon number cut-off up to 230. The Fock-state description is convenient, as the Kraus operators describing the channel, Eq.\ \eqref{eq:Kraus}, is represented in terms of annihilation operators, which have a simple and sparse Fock state representation, 
\begin{equation}
    \hat{a}=\sum_k \sqrt{k}\ket{k-1}\bra{k}.
\end{equation}
Additionally, the displacement operator can be computed as
\begin{equation}
    \hat{D}(\alpha) = \sum_{k,l} \sqrt{\frac{k!}{l!}}\alpha^{l-k}e^{-|\alpha^2/2}L_k^{l-k}(|\alpha|^2)\ket{l}\bra{k}.
\end{equation}
where $L$ are the generalized Laguerre polynomials. Finally, squeezed states are represented by
\begin{equation}
    \hat{S}(r)\ket{\textrm{vac}} = \frac{1}{\sqrt{\cosh(r)}}\sum_k\frac{\sqrt{(2k)!}}{2^k k!}(-\textrm{tanh}(r))^k\ket{2k}.
\end{equation}
This allow us to simulate the total circuit of Fig.\ \ref{fig:concept}b and compute the effective squeezing levels of the output state.\\ 
For the cat breeding protocol, the Fock representation becomes less efficient, as it involves the combination of two modes, thus squaring the required dimensionality in the Fock basis. Instead, the $p$-quadrature basis is convenient. The transformation from the Fock basis to the $p$ basis is done via the relation
\begin{equation}
    \ket{n}=\int dp\sqrt{\frac{1}{\sqrt{\pi}2^n n!}}e^{-p^2/2}H_n(p)\ket{p} \label{eq:basis}
\end{equation}
where $H_n$ is the $n$'th Hermite polynomial.
An arbitrary mixed single-mode state is described in the $p$ basis by a wavefunction $\psi(p,p')$:
\begin{equation}
    \rho = \int\int dpdp' \psi(p,p')\ket{p}\bra{p'}.
\end{equation}
Two modes with identical wavefunctions mixing on a 50:50 beamsplitter is described by the transformation:
\begin{align}
    &\psi(p_1,p_1')\psi(p_2,p_2') \nonumber\\
    &\rightarrow \psi\left(\frac{p_1+p_2}{\sqrt{2}},\frac{p_1' + p_2'}{\sqrt{2}}\right)\psi\left(\frac{p_1-p_2}{\sqrt{2}},\frac{p_1' - p_2'}{\sqrt{2}}\right).
\end{align}
Measuring mode 2 at $p_2=0$ leaves the other mode in the state
\begin{equation}
    \rightarrow \psi\left(\frac{p_1}{\sqrt{2}},\frac{p_1'}{\sqrt{2}}\right)^2.
\end{equation}
Iterating this procedure $M$ times results in the transformation 
\begin{equation}
    \psi(p,p')\rightarrow \psi\left(\frac{p}{\sqrt{2}^M},\frac{p'}{\sqrt{2}^M}\right)^{2^M}. \label{eq:p1}
\end{equation}
The expectation values used to calculate the effective squeezing levels can be evaluated as:
\begin{align}
    \langle\hat{D}(\sqrt{2\pi})\rangle&=\textrm{Tr}\left(\int dp \int dp' \psi(p,p')\hat{D}(\sqrt{2\pi})\ket{p}\bra{p'}\right)\nonumber\\
    &=\textrm{Tr}\left(\int dp \int dp' \psi(p,p')e^{-i2\sqrt{\pi}p}\ket{p}\bra{p'}\right)\nonumber\\
    &=\int dp \psi(p,p)e^{-i2\sqrt{\pi}p}, \label{eq:p2}
\end{align}
and
\begin{align}
    \langle\hat{D}(i\sqrt{2\pi})\rangle&=\textrm{Tr}\left(\int dp \int dp' \psi(p,p')\hat{D}(i\sqrt{2\pi})\ket{p}\bra{p'}\right)\nonumber\\
    &=\textrm{Tr}\left(\int dp \int dp' \psi(p,p')\ket{p+2\sqrt{\pi}}\bra{p'}\right)\nonumber\\
    &=\int dp \psi(p-2\sqrt{\pi},p). \label{eq:p3}
\end{align}
We thus compute the input cat state in the Fock-basis, then use Eq.\ \eqref{eq:basis} to calculate the mixed state wavefuncion along the lines $\left(p\sqrt{2}^M,p\sqrt{2}^M\right)$ and $\left((p-2\sqrt{\pi})\sqrt{2}^M,p\sqrt{2}^M\right)$ from which we obtain the expectation values to compute the effective squeezing levels using Eqs.\ \eqref{eq:p1}-\eqref{eq:p3}.


\section*{S7: Generation of squeezed vacuum states}

Here we show how to generate the required initial squeezed states, using the cavity QED system and coherent state inputs. The idea is that a squeezed state can be represented as a superposition of coherent states \cite{hastrup2021uncoditional}:
\begin{equation}
    \hat{S}(r)\ket{\textrm{vac}} \propto \int \exp\left(-\frac{\alpha^2}{e^{2r}-1}\right)\hat{D}(i\alpha)\ket{\textrm{vac}}d\alpha, \label{eq:SqueezedState}
\end{equation}

\begin{figure}
    \centering
    \includegraphics{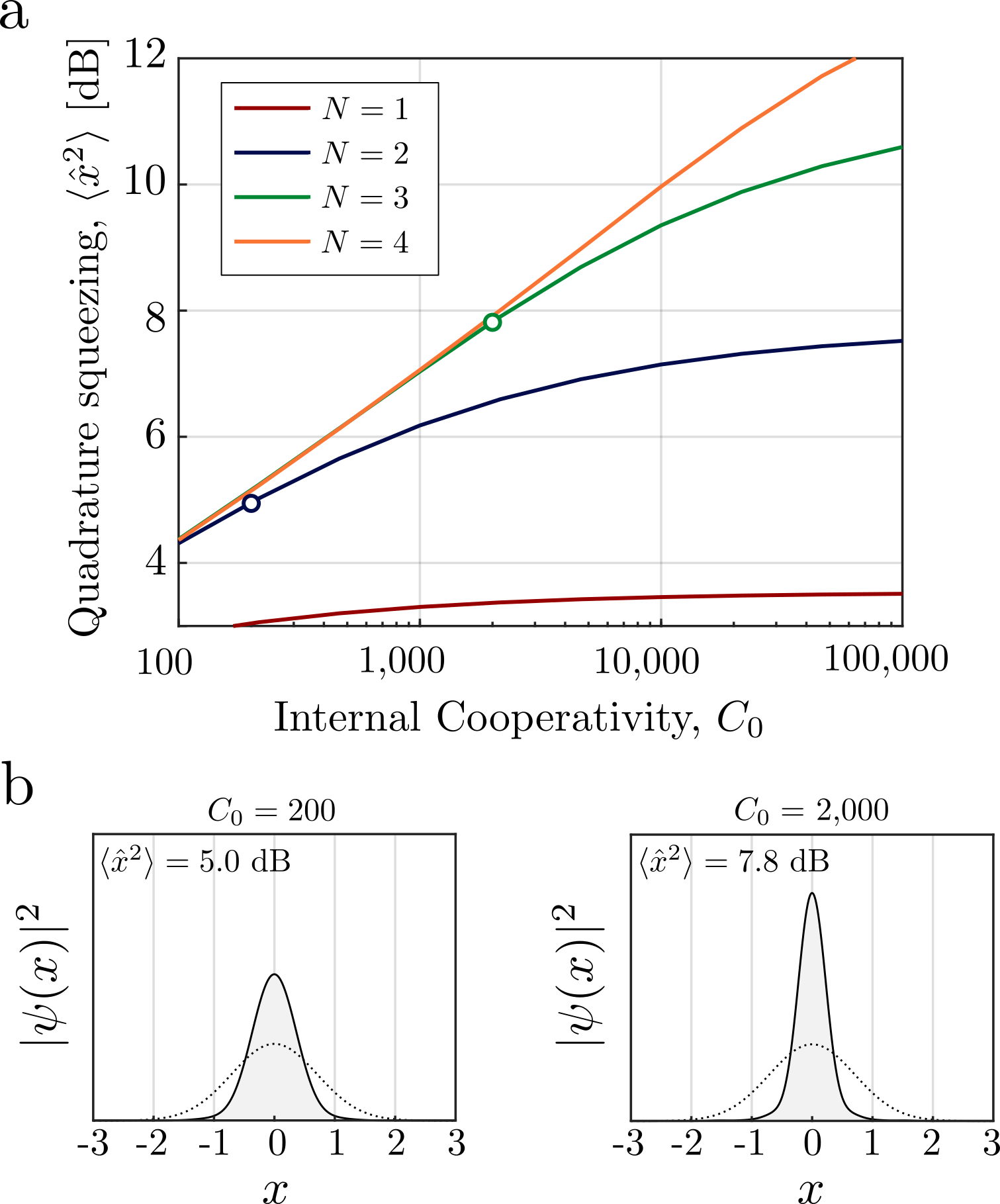}
    \caption{(a) Achievable amounts of quadrature squeezing using the protocol of Fig.\ \ref{fig:squeezing} with input vacuum states and small displacement amplitudes. (b) Example quadrature distributions of the obtainable squeezed states generated with $C_0=200$ and $C_0=2000$ choosing $N=2$ and $N=3$ respectively, corresponding to the points marked with open circles in (a). The dotted line shows the quadrature distribution of the vacuum state for comparison.}
    \label{fig:squeezing}
\end{figure}
\noindent which holds for $r>0$ and the integral is over real $\alpha$. By creating a discrete superposition of closely spaced coherent states on a line in phase space, we can approximate Eq.\ \eqref{eq:SqueezedState} to achieve an approximate squeezed state. Indeed, quadrature squeezed states were observed in \cite{hacker2019deterministic} using only two coherent states. We therefore propose to use the method of Fig.\ \ref{fig:concept}, but with a vacuum input and smaller displacement amplitudes chosen in the $p$ direction. The resulting amount of quadrature squeezing is shown in Fig.\ \ref{fig:squeezing}a, where we have numerically tuned the displacement amplitude to optimize the squeezing. As expected, as the intrinsic cooperativity of the cavity increases, we can achieve larger amounts of quadrature squeezing using a suitable number of interactions. Fig. \ref{fig:squeezing}b shows the resulting $x$ quadrature distributions at a $C_0=200$ and $C_0=2000$ for $N=2$ and $N=3$ interactions, i.e. states composed of $4$ and $8$ coherent states on the $x=0$ axis respectively, showing clear quadrature squeezing relative to the vacuum (dotted line). 

While squeezed states can be generated directly from vacuum states using the cavity QED system, comparing the results of Figs.\ \ref{fig:squeezing} and \ref{fig:breeding} the squeezing levels are lower than the approximate squeezing levels achievable using the same intrinsic cooperativity with the cat breeding method of Fig.\ \ref{fig:breeding}. Thus for the cat breeding approach without an external squeezed vacuum source, the approximate squeezing levels will be limited to those presented in Fig.\ \ref{fig:squeezing}. 
\end{widetext}


\end{document}